\documentclass[lettersize,journal]{IEEEtran}
\usepackage{amsmath,amsfonts}
\usepackage{algorithmic}
\usepackage{array}
\usepackage[caption=false,font=normalsize,labelfont=sf,textfont=sf]{subfig}
\usepackage{textcomp}
\usepackage{stfloats}
\usepackage{url}
\usepackage{verbatim}
\usepackage{graphicx}
\usepackage{multirow}
\usepackage{cite}
\usepackage{cuted}
\usepackage{authblk}
\def\BibTeX{{\rm B\kern-.05em{\sc i\kern-.025em b}\kern-.08em
    T\kern-.1667em\lower.7ex\hbox{E}\kern-.125emX}}
\usepackage{balance}
\begin{document}
\title{Beacon-enabled TDMA Ultraviolet Communication Network System Design and Realization}
\author[1]{Yuchen Pan}
\author[1]{Fei Long}
\author[2]{Ping Li}
\author[2]{Haotian Shi}
\author[2]{Jiazhao Shi}
\author[2]{Hanlin Xiao}
\author[1,*]{Chen Gong}
\author[1]{Zhengyuan Xu}

\affil[1]{key Laboratory of Wireless-Optical Communications, Chinese Academy of Sciences, School of Information Science and Technology, University of Science and Technology of China, Hefei, Anhui 230026, China}

\affil[2]{Xi'an Modern Control Technology Research Institute, Xi'an 710065}

\affil[*]{Corresponding author: cgong821@ustc.edu.cn}

\markboth{Journal of \LaTeX\ Class Files,~Vol.~18, No.~9, September~2020}%
{How to Use the IEEEtran \LaTeX \ Templates}

\maketitle

\begin{abstract}
	Nonline of sight (NLOS) ultraviolet (UV) scattering communication can serve as a good candidate for outdoor optical wireless communication (OWC) in the cases of non-perfect transmitter-receiver alignment and radio silence. We design and demonstrate a NLOS UV scattering communication network system in this paper, where a beacon-enabled time division multiple access (TDMA) scheme is adopted. In our system, LED and PMT are employed for transmitter and receiver devices, repectivey. Furthermore, we design algorithms for beacon transmission, beacon reception, time compensation, and time slot transition for hardware realization in field-programmable gate array (FPGA) board based on master-slave structure, where master node periodically transmits beacon signals to slave nodes. Experimental results are provided to evaluate the time synchronization error and specify the system key parameters for real-time implementation. We perform field tests for real-time communication network with the transmission range over $110 \times 90$m$^{2}$, where the system throughput reaches $800$kbps.
\end{abstract}


\section{Introduction}
Non-line of sight (NLOS) scattering communication can offer optical wireless communication link for radio-silence scenarios while not requiring perfect alignment of the transmitter and receiver \cite{1,2,3,8}. Existing works based on NLOS visble light communication (VLC) have been reported in \cite{4,5,6}. However, NLOS VLC suffers from strong solar interference especially during daytime. Compared with VLC, NLOS ultraviolet (UV) communication is adopted in outdoor environments during daytime since the background radiation of UV spectrum within spectrum between $200$nm and $280$nm is absorbed by the atmosphere \cite{2}. Hence, the characteristics of UV \cite{7} leads to prospective application of UV communication in the case of non-perfect transmitter-receiver alignment and strong background radiation. 

NLOS UV scattering communication has been extensively investigated from theoretic and experimental perspectives. Different from RF-based communication, the received signal of UV communication exhibits the characteristics of dicrete photoelectrons due to the extreme path loss. Hence, the Poisson-type channel has been adopted for investigating UV channel capacity and achievable rate. The capacities of Poisson channel have been studied in \cite{information1,information2,information3,information4,information5,information6,information7}. The capacities of continuous-time and discrete-time Poisson channel have been investigated in \cite{information1,information2} and \cite{information3,information4}, respectively. The capacity of MISO has been reported in \cite{information5}. Moreover, the capacities of Poisson fading channel and MIMO Poisson fading channel have been studied in \cite{information7} and \cite{information6}, respectively. Regarding UV channel link gain, existing works based on the UV channel model have been reported in \cite{lg1,lg2,lg3,lg4}, which have investigated the channel link gain from Monte Carlo, theoretic and experimental pespective. Futhermore, the characterization and detection of the UV receiver-side signal have been studied in \cite{cha1,cha2,cha3,cha4}. Channel estimation of UV channel parameters have been investigated in \cite{cha2,cha4}, while the characterization on practical photon counting receiver has been discussed in \cite{cha1} and the signal detection for optical scattering communication in different scenarios have been studied in \cite{cha2,cha3}. Other physical layer techniques have been reported in \cite{phy1,phy2}, including polar codes in wireless communications \cite{phy1} and spectrum sensing for optical wireless scattering communications \cite{phy2}. Besides the theoretical investigation, experimental works on UV scattering communications have beem discussed in \cite{ex1,ex2,ex3,ex4}.

UV communication networks are widely investigated for a network scenario to enhance system performance. The Poisson multiple access channel has been investigated in \cite{netinfo1}. The connectivity performance for UV networks has been studied in \cite{connect1}. Existing works based on the medium-access control (MAC) techniques for UV communication networks have been reported in \cite{MAC1,MAC4}, while signal characterization and detection for multiple access NLOS communication have been investigated in \cite{MAC2,MAC3}.

In this paper, we realize the real-time UV time division multiple access (TDMA) communication network, consisting one master node and multiple slave nodes, over $90 \times 100m^{2}$ with over $800$ kbps throughput via applying beacon-based time sychronization. We employ a UV LED as the transmitter and photomultiplier tubes (PMTs) as the receiver. We design and realize signal processing algorithms based on the discrete-time Poisson channel model, consisting of beacon transmission, beacon reception, time compensation and time slot transition. Experiments are conducted to measure the time sychronization error between nodes such that the guard intervals can be designed to avoid collision. Furthermore, we realize a real-time $4$-node UV communication network system consisting of multiple PMTs, field-programmable gate array (FPGA) boards for implementation of digital signal processing. We complete a real-time UV network communication experiment with the transmission range over $90 \times 100 m^{2}$  and the resulting throughput can reach $800$ kbps.

This remainder is organized as follows. In Section \ref{block}, we provide the block diagrams of the proposed communication network system. In Section \ref{process}, we address the signal processing of beacon-based time synchronization and time slot transition. Experimental results are shown in Section \ref{system design}. Finally, we conclude this paper in Section \ref{conclusion}.


\section{Signal Characterization and System Block Diagram}\label{block}
\subsection{Signal Characterization of UV Communication}
We adopt UV spectrum for communication due to its scattering property to enhance system reception range and limited transmission distance to enhance security. The received signals at the receiver side are in the discrete-photon level in the form of photoelectrons, whose number satisfies Poisson distribution. Let $N_i$ denote the number of received photoelectrons from node $i$ within symbol duration and $S$ denote the transmitted symbol. $N_{i}$ satisfies the following Poisson distribution:
\begin{align}
	P(N_i = n | S = 1) &= {\dfrac{{\lambda_{s,i}+\lambda_{b,i}}^n}{n!}}e^{{\lambda_{s,i}+\lambda_{b,i}}},\\
	P(N_i = n | S = 0) &= {\dfrac{{\lambda_{s,i}}^n}{n!}}e^{{\lambda_{s,i}+\lambda_{b,i}}},
\end{align}
where $\lambda_{s,i}$, $\lambda_{b,i}$ denote the mean number of received photoelectrons at node $i$ for the signal and background components.

\subsection{Wireless Communication Network System Block Diagram}
\begin{figure}[htbp]
	\centering
	\includegraphics[width=0.4\textwidth]{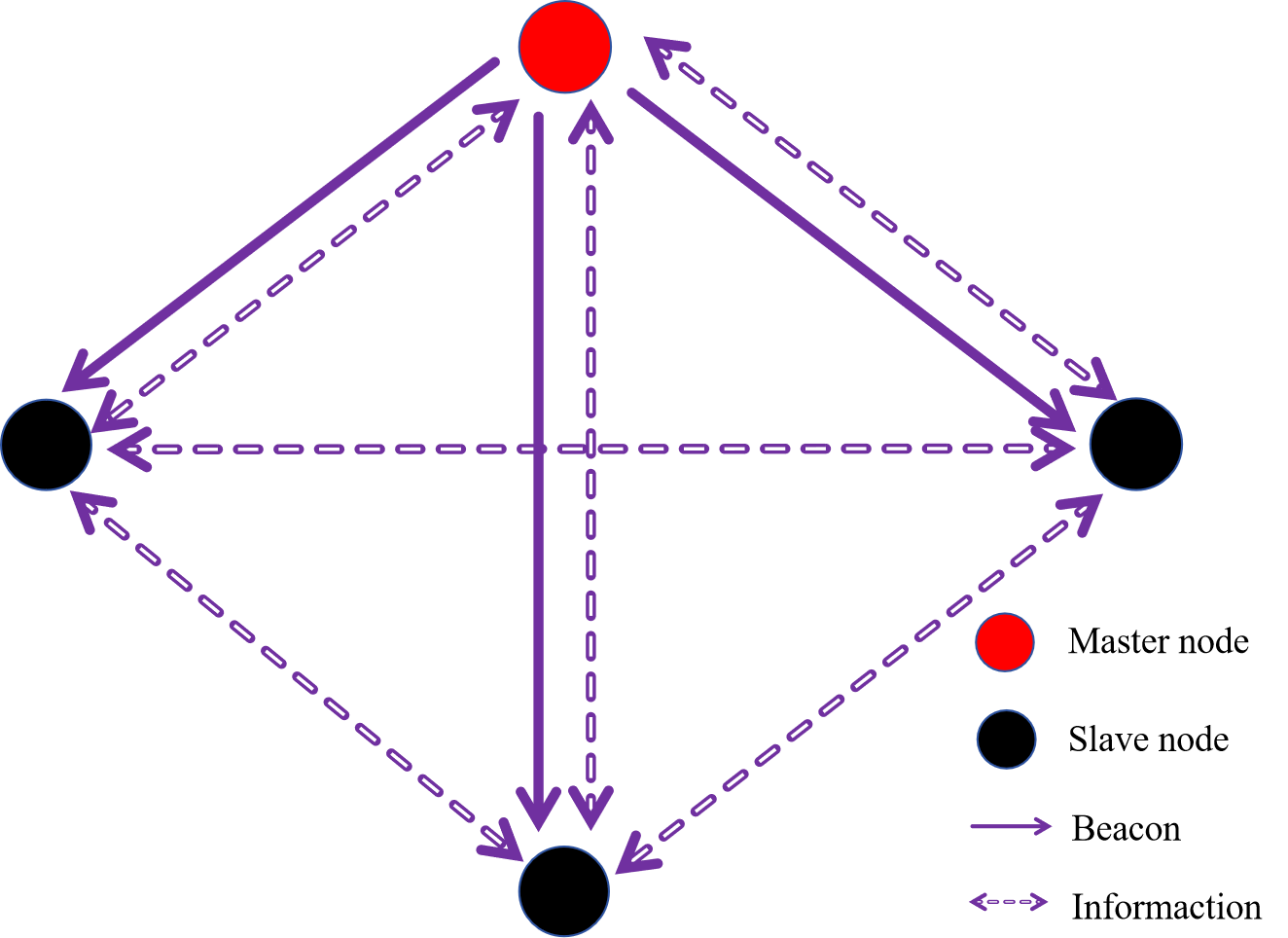}
	\caption{Example of a network topology.}
	\label{topology}
\end{figure}

We consider a $N$-node network, consisting of one master node (denoted as node $1$) and ($N-1$) slave nodes (denoted as node $2,3,\cdots N$, respectively). Each node works as a transceiver. Any two nodes can communicate with each other through one hop. TDMA MAC scheme is adopted. The master node broadcasts beacon signal to the slave nodes. Each node accesses the channel according to the allocated slot after time sychronization is performed by the slave nodes based on the beacon signal reception. An example of $4$-node network is shown in Fig. \ref{topology} to show the topology and the data flow of the network system.

The FPGAs are adopted as the data processing unit of each node. The information bits are randomly generated in the personal computer(PC). The information bits are divied into groups of $8$ bits and sent from the PC to the FPGA board via UART interface. In addition, the beacon signals are generated in the FPGA board of the master node. LEDs are employed as the UV transmitter and we adopt OOK modulation. The information bits and beacon signals are exported to modulate the UV LED via the pin of the FPGA board.

\begin{figure*}[htp]
	\centering
	\includegraphics[width=0.9\textwidth]{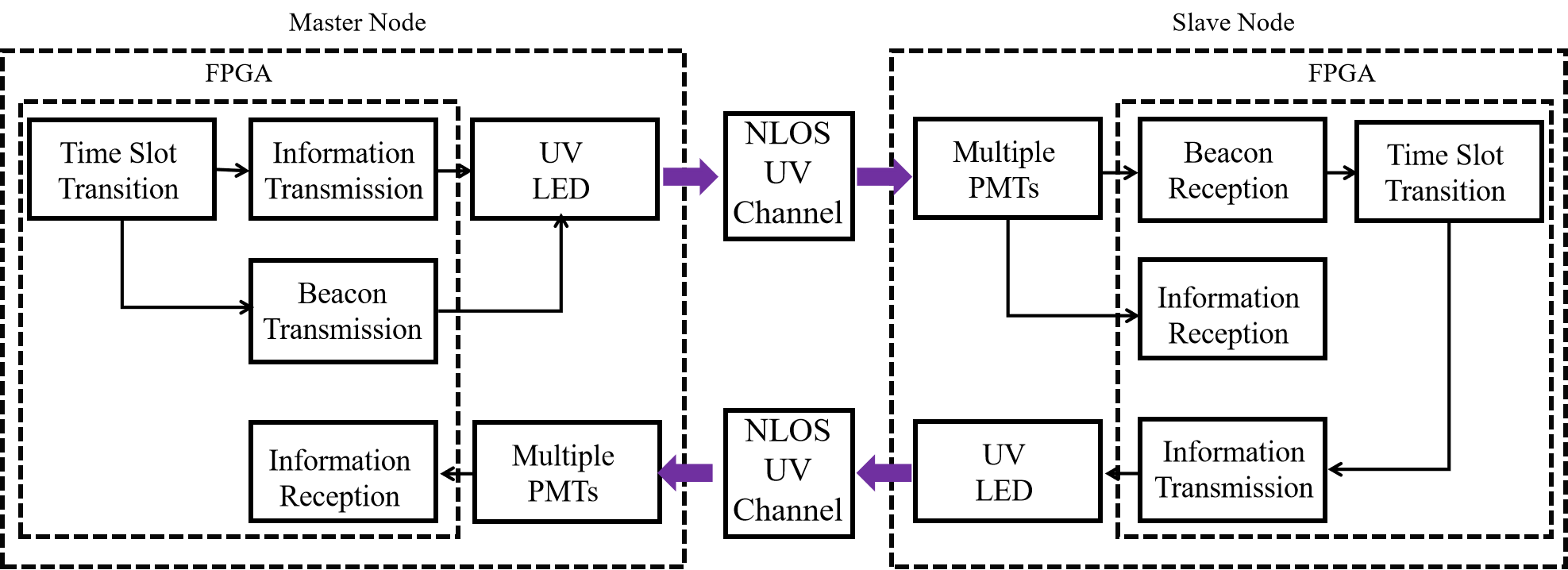}
	\caption{The block diagram of NLOS UV Network System.}
	\label{Block}
\end{figure*}

\begin{figure*}[htp]
	\centering
	\includegraphics[width=0.9\textwidth]{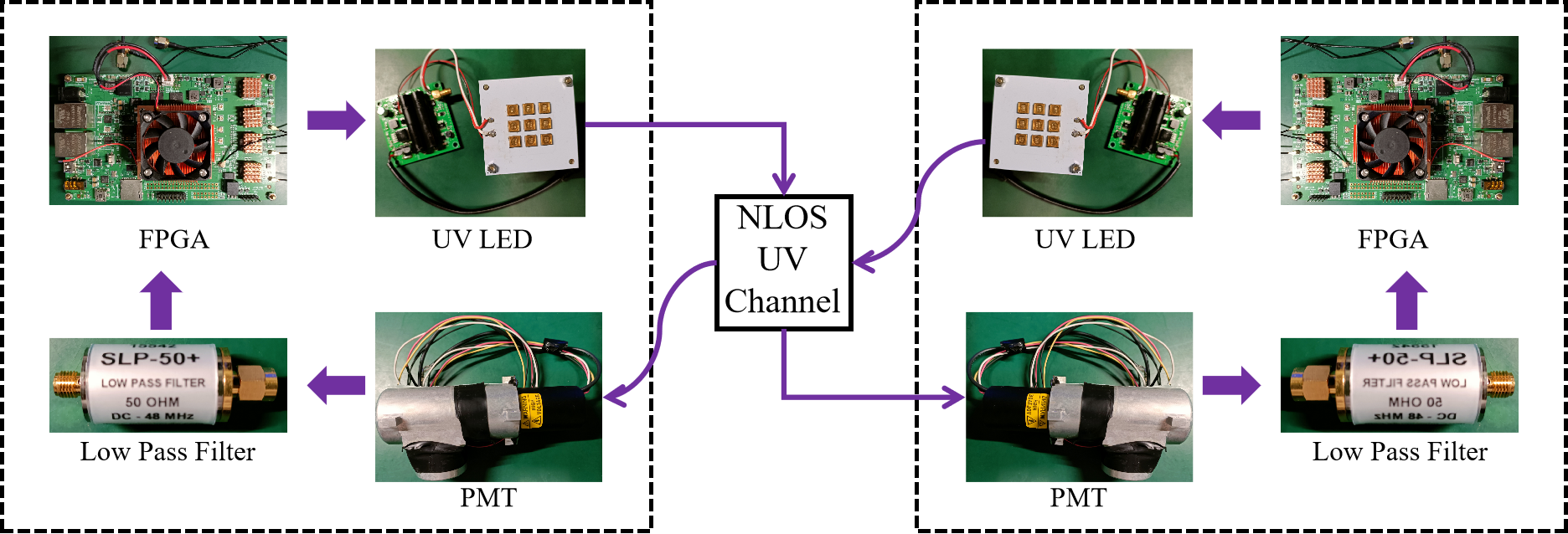}
	\caption{The hardware realization blocks of the NLOS scattering communication network system.}
	\label{hardware}
\end{figure*}

PMTs are employed as the photon-detector, which can convert the received photoelectrons to analog pulses such that photon counting can be realized via pulse counting. A PMT(CR340) detector is sealed into a shelding box, which is integrated with a UV optical filter that passes the light signal of wavelenth around $266$ nm and blocks signals on other wavelenths. The digital voltage values are achieved via analog-to-digital converters(ADCs). The digital processing of beacon reception, time compensation, time slot transition is performed in the FPGA boards of the slave nodes and the digital processing of frame sycnhronization, channel estimation and symbol detection are performed in the FPGA boards of all nodes. The block diagram and experimental system of the NLOS UV scattering communication network are shown in Figs. \ref{Block} and \ref{hardware}, respectively.

\section{Digital Processing in Hardware Realization}\label{process}

\subsection{Beacon Transmission }\label{transmission}
The master node automatically transmits periodic beacon signal with period $T$ to the slave nodes after FPGA board of the master node working. We adopt $L$-bit binary $m$ sequence as the beacon signal due to its excellent autocorrelation.

The time counter of the master node will start working as it starts transmitting beacon signal. Let $C_{i}$ denote the time counter of node $i$. $C_{1}$ counts from $0$ to $(C_{max} - 1)$ periodically, where $C_{max} \times t_{clock} = T$ and $t_{clock}$ denotes the interval of two consecutive clock rising edge.

The time synchronization pulses are generated by the slave nodes via finding the start of beacon signal such that the slave nodes can achieve time synchronization as the master node. We divide the received signals into chips with the duration $T_c = \dfrac{T_s}{M}$, where $T_s$ denotes the symbol duration and $M$ denotes the number of chips in one symbol. The counting-based synchronization in \cite{Guan} is adopted for beacon reception. The position with the maximum correlation peak corresponds to the start of beacon signal. 
\subsection{Time Compensation and Time Slot Transition}
\subsubsection{Time Delay Analysis}\label{time delay}
\begin{figure}[htbp]
	\centering
	\includegraphics[width=0.45\textwidth]{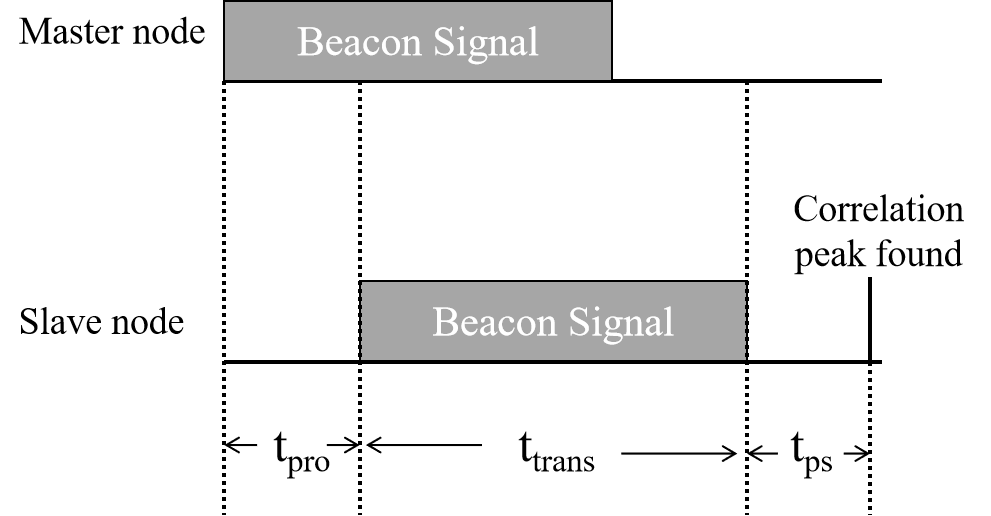}
	\caption{Demonstration of three time delays.}
	\label{relationship}
\end{figure}

Based on the details in \ref{transmission}, the time delay between beacon transmission and time sychronization pulse generation consists of transmission delay $t_{trans}$, propagation delay $t_{pro}$ and processing delay $t_{ps}$, as shown in Fig. \ref{relationship}. $t_{trans}$ and $t_{pro}$ are defined as in:
\begin{align}
	t_{trans} &= {L \times T_s},\\
	t_{pro}^{ij} &= \dfrac{r_{ij}}{C},1 \leq i,j \leq N,i \neq j,
\end{align}
where $t_{pro}^{ij}$ denotes propagation between node $i$ and $j$, $C$ denotes the speed of light, and $r_{ij}$ denotes the distance between node $i$ and $j$, respectively. The processing delay should be experimentally measured since it is a random variable and is difficult to analyze via mathematical model. Let $t_{ps}^{i}$ denote the processing delay of node $i$. Without loss of generality, assume that $t_{ps}^{i}$,$1 \leq i \leq N$ are independent and identically distributed.

\subsubsection{Time Compensation}
The time counters of the slave nodes start to count as the time synchronization pulses are achieved by the slave nodes. If the time counters of the slave nodes count from $0$, the time delay shown in \ref{time delay} will still exist. The time compensation should be performed at the slave nodes due to the time delay. Instead of $0$, $C_{i},i=2,3,\cdots,N$ will be set to the initial value $c_{initial}$ as the time sychronization pulse is generated by node $i$. $c_{initial}$ is defined as:
\begin{equation}
	c_{initial} = \dfrac{t_{trans} + \mathop{\max}\limits_{j}{t_{pro}^{1j}} + \tilde{t_{ps}}}{t_{clock}}
\end{equation}
where $\tilde{t_{ps}}$ denotes the experimentally measured processing delay. Let $t_{d}^{i}$ denote the time synchronization error between node $i$ and the master node, $t_{d}^{i}$ is defined as :
\begin{equation}
	t_{d}^{i} = \left\{
	\begin{aligned}
		& (\tilde{t_{ps}} - t_{ps}) + (\mathop{\max}\limits_{j}{t_{pro}^{1j}} - t_{pro}^{1i}) &,& 2 \leq i \leq N, \\
		& 0 &,&i=1.
	\end{aligned}
	\right.
\end{equation}	

\subsubsection{Time Slot Transition}
\begin{figure}[h]
	\centering  
	
	\subfloat[Master node]{
		\includegraphics[width=0.4\textwidth]{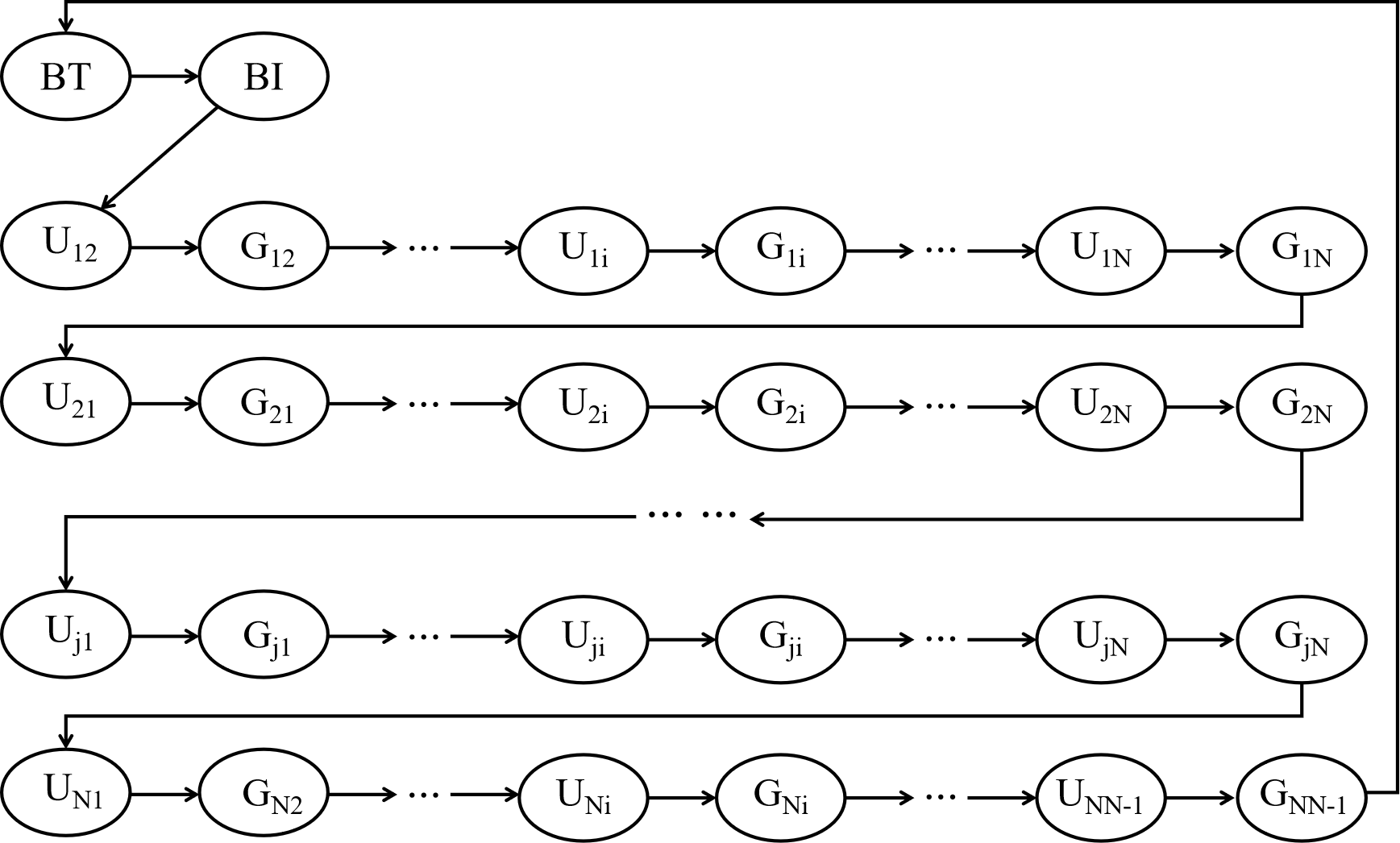}}
	\\
	\subfloat[Slave node]{
		\includegraphics[width=0.4\textwidth]{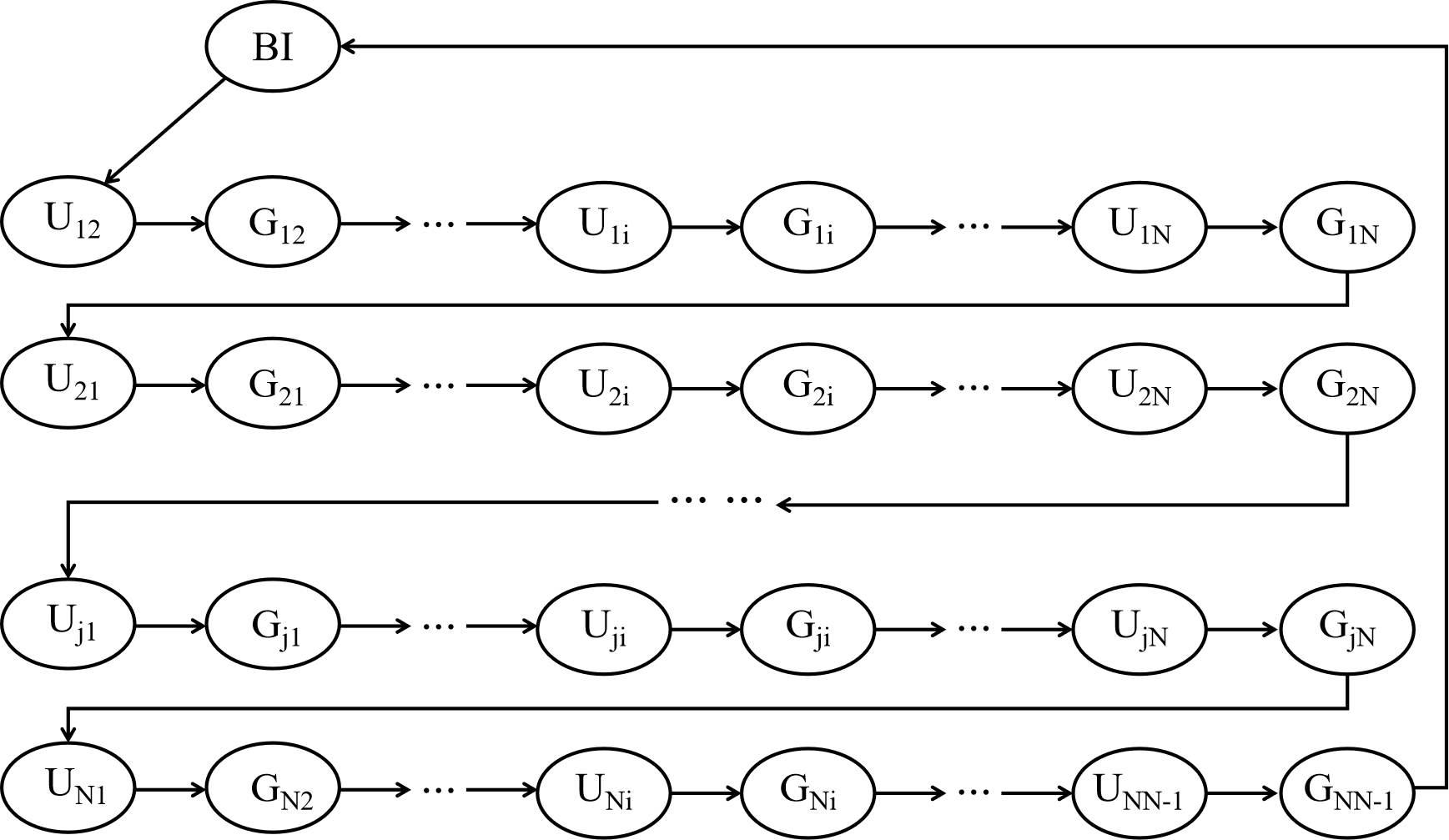}}
	\caption{State transition of the master node and the slave nodes.}
	\label{transition}
\end{figure}

Each time period is divided into multiple slots based on TDMA MAC scheme. The time period of the master node consists of $1$ beacon transmission slot, $1$ beacon interval slot, $N\times(N-1)$ information transmission slots and $N\times(N-1)$ guard interval slots. Compared with the master node, the slave nodes don't have the beacon transmission slot. The slot transition depends on the time counter of each node. The time slot transition of the master node and the slave nodes is shown in Fig. \ref{transition}. Let $BT$, $BI$ denote beacon transmission slot, beacon interval slot, respectively. The beacon transmission slot is for the master node to transmit beacon signal to the slave nodes, and the beacon interval slot is for the slave nodes to decode the beacon signal. Let $t_{bt}$ and $t_{bi}$ denote the duration of the beacon transmission slot and the duration of the beacon interval slot, respectively. $t_{bt}$ and $t_{bi}$ are subject to :
\begin{align}
	t_{bt} &= {L \times T_s},\\ \label{BI}
	t_{bi} &\geq ( \mathop{\max}\limits_{j}{t_{pro}^{1j}} + \tilde{t_{ps}}),\\ 
	t_{d}^{i} &\leq \lvert t_{bi} - ( \mathop{\max}\limits_{j}{t_{pro}^{1j}} + \tilde{t_{ps}}) \rvert.
\end{align}

\begin{figure}[htbp]
	\centering
	\includegraphics[width=0.43\textwidth]{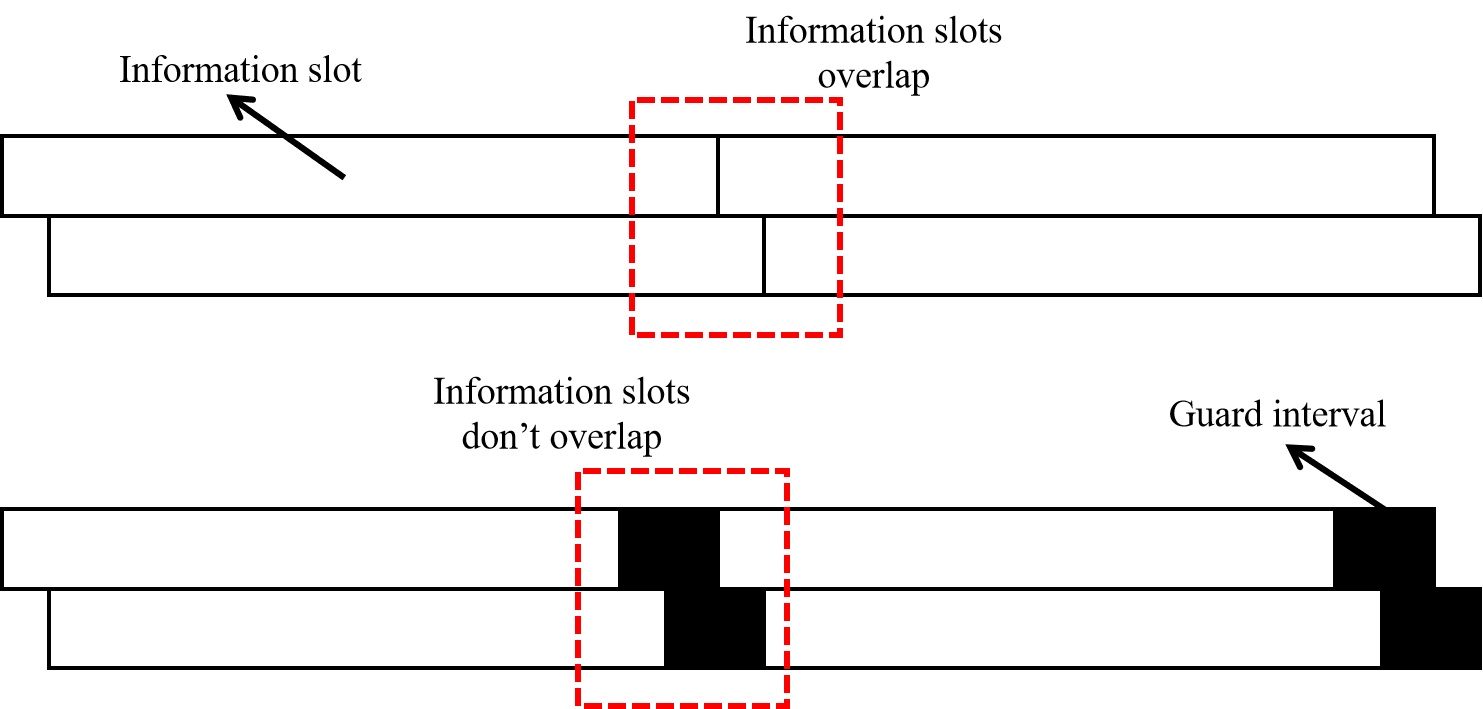}
	\caption{Comparison between "with guard interval" and "without interval".}
	\label{guard}
\end{figure}

The information slots are designed for nodes in the network to transmit information frames. $U_{ij}$ in Fig. \ref{transition} denotes the time slot for node $i$ transmitting frames to node $j$. The guard intervals are designed to avoid the collision caused by time sychronization error of different nodes. $G_{ij}$ in Fig. \ref{transition} denotes the guard interval after time slot $U_{ij}$. Without generality, all guard intervals and information slots are set as the same length. Let $t_g$ denote the duration of the guard interval and $t_g$ is subject to :
\begin{equation}\label{GI}
	t_{g} \geq \lvert t_{d}^{i} - t_{d}^{j} \rvert, \forall 1 \leq i,j \leq N. 
\end{equation}	
Fig. \ref{guard} demonstrates the necessity of the guard interval. Based on the above demonstration, the time slot transition conditions of master node and slave node $k$ are given in Equation. (\ref{master_condition})-(\ref{master_end}) and Equation. (\ref{slave_condition})-(\ref{slave_end}), respectively, where $t_{u}$ denotes the length of one information slot.

\begin{figure*}[htb]
	\begin{align}\label{master_condition}
		BT\rightarrow BI: \ &C_{1}=\frac{t_{bt}}{t_{clock}}, \\ 
		BI\rightarrow U_{12}: \ &C_{1}=\frac{t_{bt}+t_{bi}}{t_{clock}}, \\ 
		U_{ij}\rightarrow G_{ij}: \ &C_{1}=\frac{t_{bt}+t_{bi}+((i-1)(N-1)+(j-1))(t_u+t_g)+t_{u}}{t_{clock}},1\leq j\leq i-1,2\leq i\leq N, \\
		 \ &C_{1}=\frac{t_{bt}+t_{bi}+((i-1)(N-1)+(j-2))(t_u+t_g)+t_u}{t_{clock}},i+1\leq j\leq N,1\leq i\leq N-1, \\
		G_{ij}\rightarrow U_{i(j+1)}: \ &C_{1}=\frac{t_{bt}+t_{bi}+((i-1)(N-1)+j)(t_u+t_g)}{t_{clock}},1\leq j\leq i-1,2\leq i\leq N, \\
		 \ &C_{1}=\frac{t_{bt}+t_{bi}+((i-1)(N-1)+(j-1))(t_u+t_g)}{t_{clock}},i+1\leq j\leq N,1\leq i\leq N-1, \\		
		G_{i(i-1)}\rightarrow U_{i(i+1)}: \ &C_{1}=\frac{t_{bt}+t_{bi}+N(i-1)(t_u+t_g)}{t_{clock}},2\leq i\leq N-1, \\
		G_{iN}\rightarrow U_{(i+1)1}: \ &C_{1}=\frac{t_{bt}+t_{bi}+i(N-1)(t_u+t_g)}{t_{clock}},1\leq i\leq N-1,\\
		G_{N(N-1)}\rightarrow BT: \ &C_{1}=\frac{t_{bt}+t_{bi}+N(N-1)(t_u+t_g)}{t_{clock}}.\label{master_end}
	\end{align}
	\hrulefill
\end{figure*}

\begin{figure*}[htb]
	\begin{align}\label{slave_condition}
		BI\rightarrow U_{12}: \ &C_{k}=\frac{t_{bt}+t_{bi}}{t_{clock}}, \\ 
		U_{ij}\rightarrow G_{ij}: \ &C_{k}=\frac{t_{bt}+t_{bi}+((i-1)(N-1)+(j-1))(t_u+t_g)+t_{u}}{t_{clock}},1\leq j\leq i-1,2\leq i\leq N, \\
		\ &C_{k}=\frac{t_{bt}+t_{bi}+((i-1)(N-1)+(j-2))(t_u+t_g)+t_u}{t_{clock}},i+1\leq j\leq N,1\leq i\leq N-1, \\
		G_{ij}\rightarrow U_{i(j+1)}: \ &C_{k}=\frac{t_{bt}+t_{bi}+((i-1)(N-1)+j)(t_u+t_g)}{t_{clock}},1\leq j\leq i-1,2\leq i\leq N, \\
		\ &C_{k}=\frac{t_{bt}+t_{bi}+((i-1)(N-1)+(j-1))(t_u+t_g)}{t_{clock}},i+1\leq j\leq N,1\leq i\leq N-1, \\		
		G_{i(i-1)}\rightarrow U_{i(i+1)}: \ &C_{k}=\frac{t_{bt}+t_{bi}+N(i-1)(t_u+t_g)}{t_{clock}},2\leq i\leq N-1, \\
		G_{iN}\rightarrow U_{(i+1)1}: \ &C_{k}=\frac{t_{bt}+t_{bi}+i(N-1)(t_u+t_g)}{t_{clock}},1\leq i\leq N-1,\\
		G_{N(N-1)}\rightarrow BI: \ &\text{The beacon signal of new period is decoded}.\label{slave_end}
	\end{align}
	\hrulefill
\end{figure*}

\section{System Design with Experimental Results}\label{system design}
\subsection{System Specification with Experimental Results}\label{specification}

We specify the NLOS UV communication system under consideration. Let the number of beacon signal bits $L=256$, the transmission symbol rate of OOK modulation $R_{s} = \frac{1}{T_{s}} = 2$Mbps, the number of chips within a symbol duration $M=10$, and the time sychronization period $T$=1s. According to \ref{BI} and \ref{GI}, the beacon interval and guard intervals are set to avoid the time slot overlap between different nodes. Hence, the time sychronizaiton error test is conducted to specify the length of the beacon interval and guard intervals. 

\begin{figure}[htbp]
	\centering
	\includegraphics[width=0.43\textwidth]{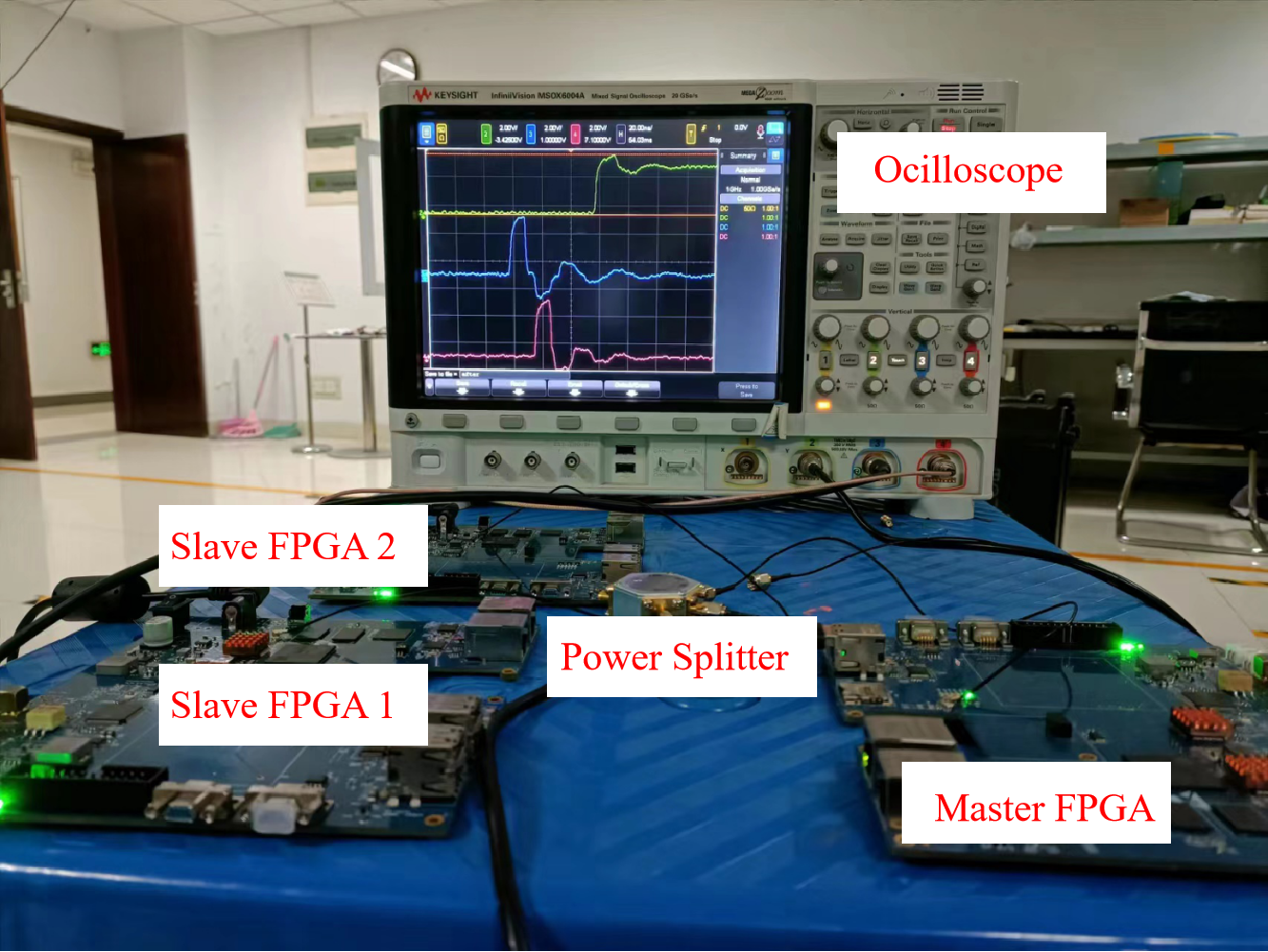}
	\caption{The experimental scenario of time synchronization error test.}
	\label{time_scenario}
\end{figure}

We conduct real-time lab test on the time synchronizaiton error according to the scenario shown in Fig. \ref{time_scenario}. A start pulse $P_{m}$ is generated by the FPGA board of the master node as it transmits the beacon signals. Synchronization pulses $P_{s}$ are generated by the FPGA boards of the slave nodes as the slave nodes receive the beacon signals successfully. We can observe the time difference of the nodes via observing the distance of the pulses on the oscilloscope. 

\begin{figure}[htbp]
	\centering
	\includegraphics[width=0.43\textwidth]{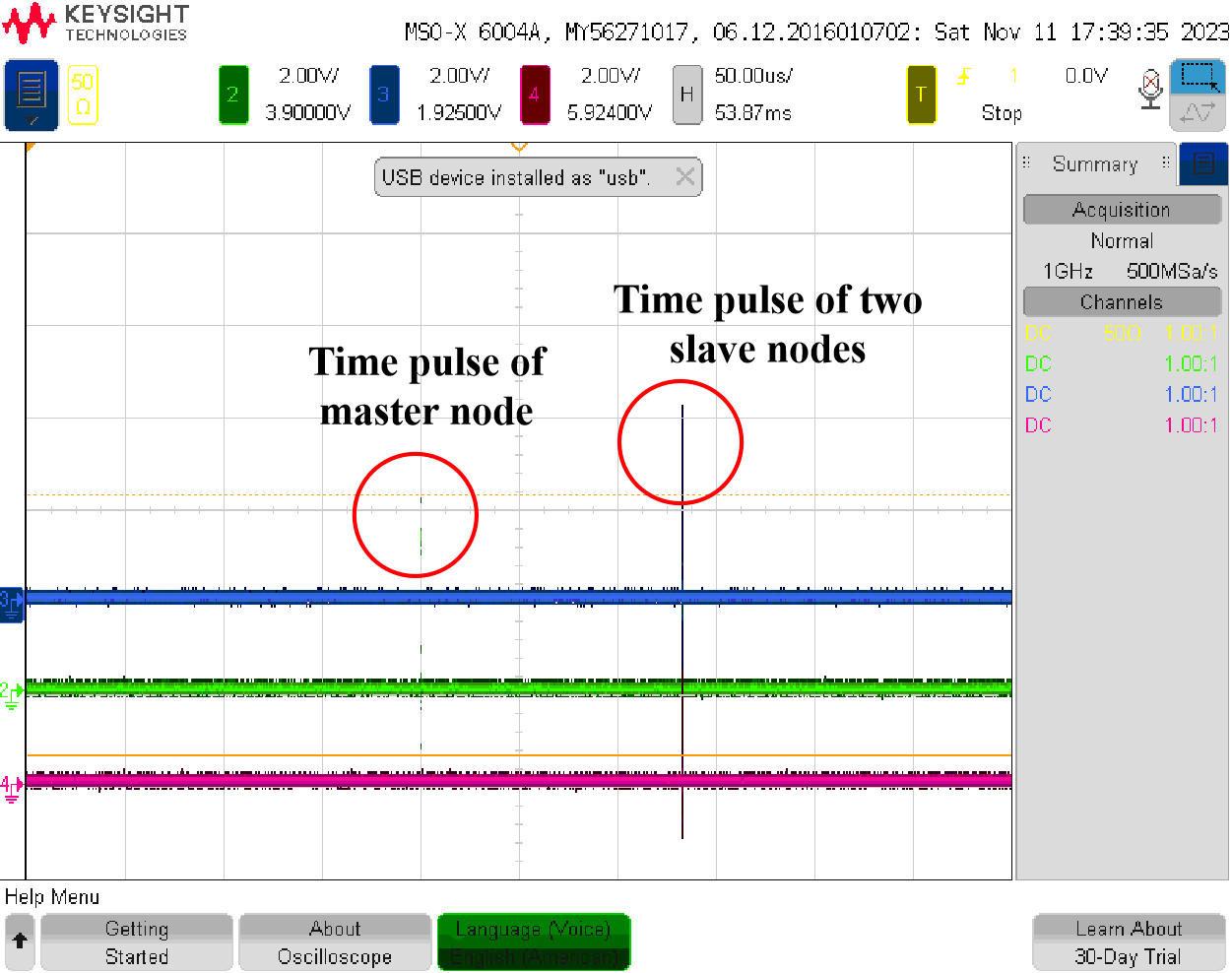}
	\caption{Time synchronization error between nodes.}
	\label{error}
\end{figure}

\begin{figure}[htbp]
	\centering
	\includegraphics[width=0.43\textwidth]{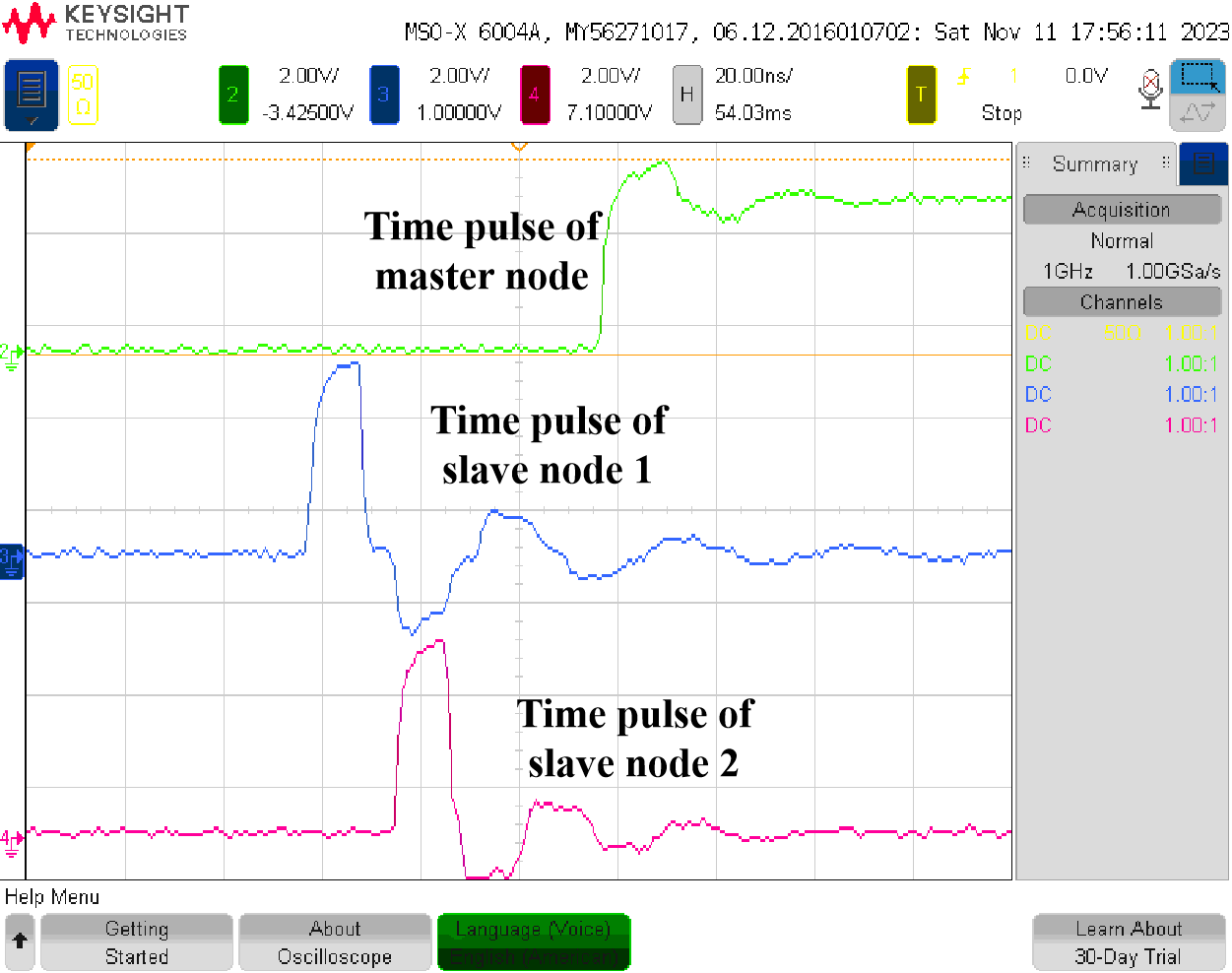}
	\caption{Time synchronization error between nodes after time compensation.}
	\label{error_compensation}
\end{figure}

\begin{figure}[h]
	\centering  
	\subfloat[Error between master node and slave node]{
		\includegraphics[width=0.4\textwidth]{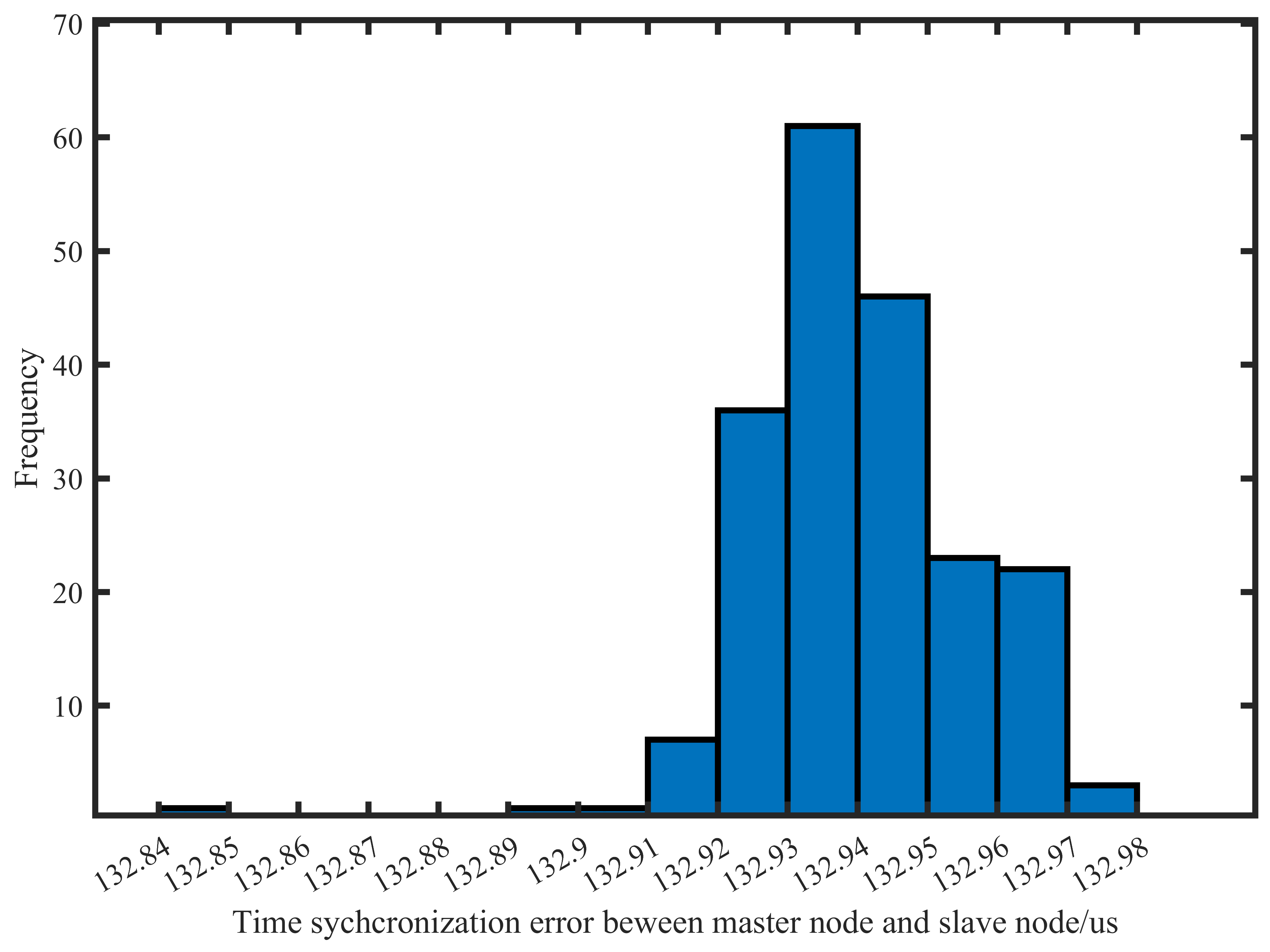}}
	\\
	\subfloat[Error between slave nodes]{
		\includegraphics[width=0.4\textwidth]{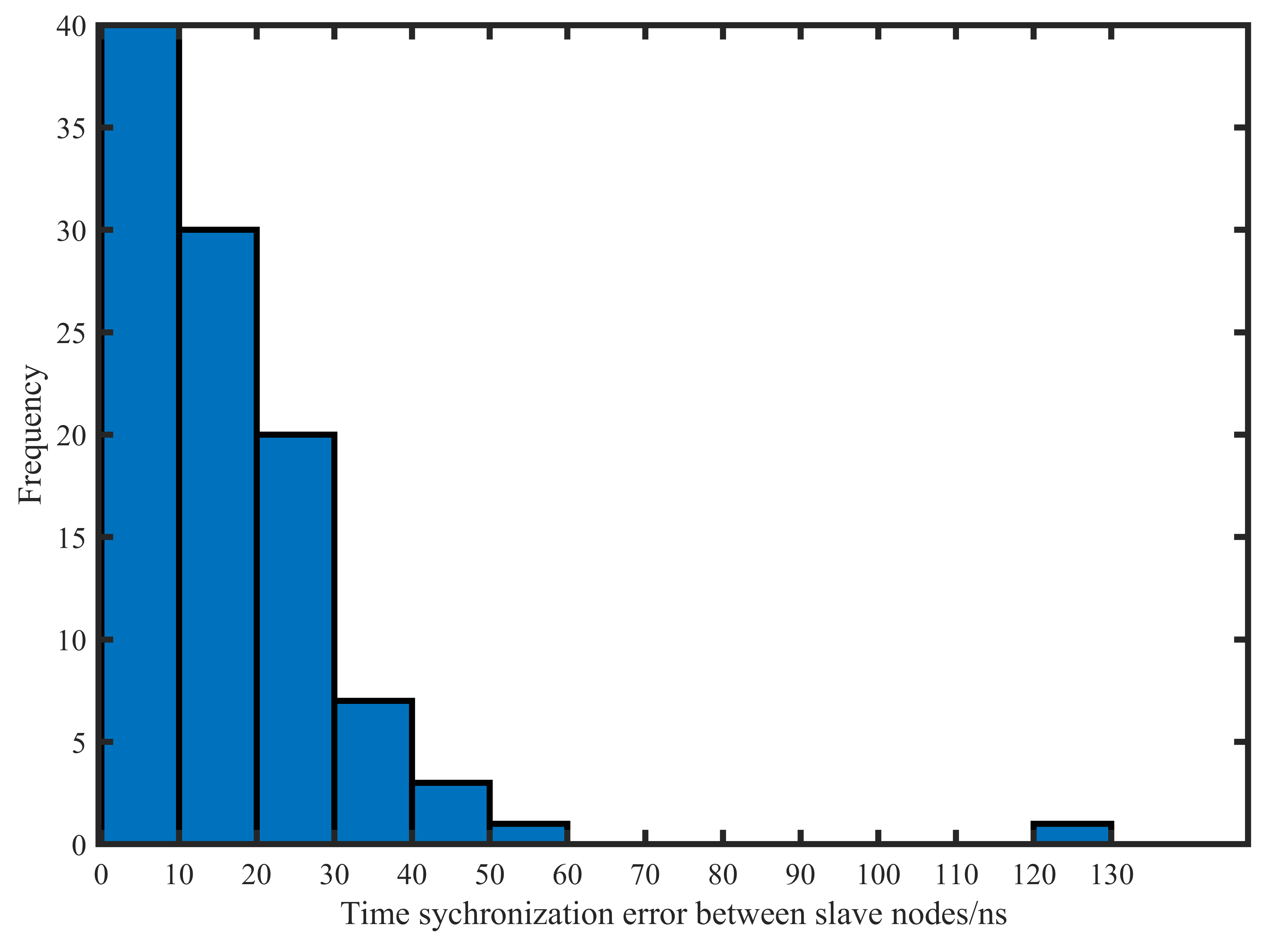}}
	\caption{Empirical probability distribution of time synchronization.}
	\label{distribution}
\end{figure}

We show the time sychronization error in Fig. \ref{error}. The time pulses of two slave nodes overlaps due to the display scale of the oscilloscope. We adopt the independent repeated trials to achieve the distribution of the time synchronization error. We measure the time sychronization error $100$ times and the distribution of the time synchronization error is shown in Fig. \ref{distribution}. The statistical variance and expectation of the time sychronization error are $0.015$ and $132.941$us, respectively. Moreover, it can be observed from the measured data that the maximum value of the data is less than $133$us. Hence, $c_{initial}$ is set as $1.33e-4$s to avoid the time collision between the master node and the slave nodes. The results after time compensation are shown in Fig. \ref{error_compensation}. The green pulse, blue pulse and red pulse denote the time of the master node, slave node $1$ and slave node $2$, respectively. It can be observed that the time synchronization errors between master node and slave nodes drops from $100$-microsecond order to $50$-nanosecond order. We consider $t_{bi}$ to be set to $128$us for processing, which is subject to Eq. \eqref{BI}. The length of all types of time slots is shown in Table \ref{slot_len}.

\begin{table}[htbp]
	\begin{center}
		\caption{The Length of All Types of Time Slots}
		\label{slot_len}
		\begin{tabular}{| c | c |}
			\hline
			Type of  time slot& Length/symbol \\
			\hline
			Beacon transmission& $256$\\
			\hline
			Beacon interval&$256$ \\ 
			\hline
			Information transmission& $137500$\\
			\hline 
			Guard interval& $29124$\\
			\hline
		\end{tabular}
	\end{center}
\end{table}

\subsection{Lab Test on the Communication Network Performance}

\begin{figure}[htbp]
	\centering
	\includegraphics[width=0.4\textwidth]{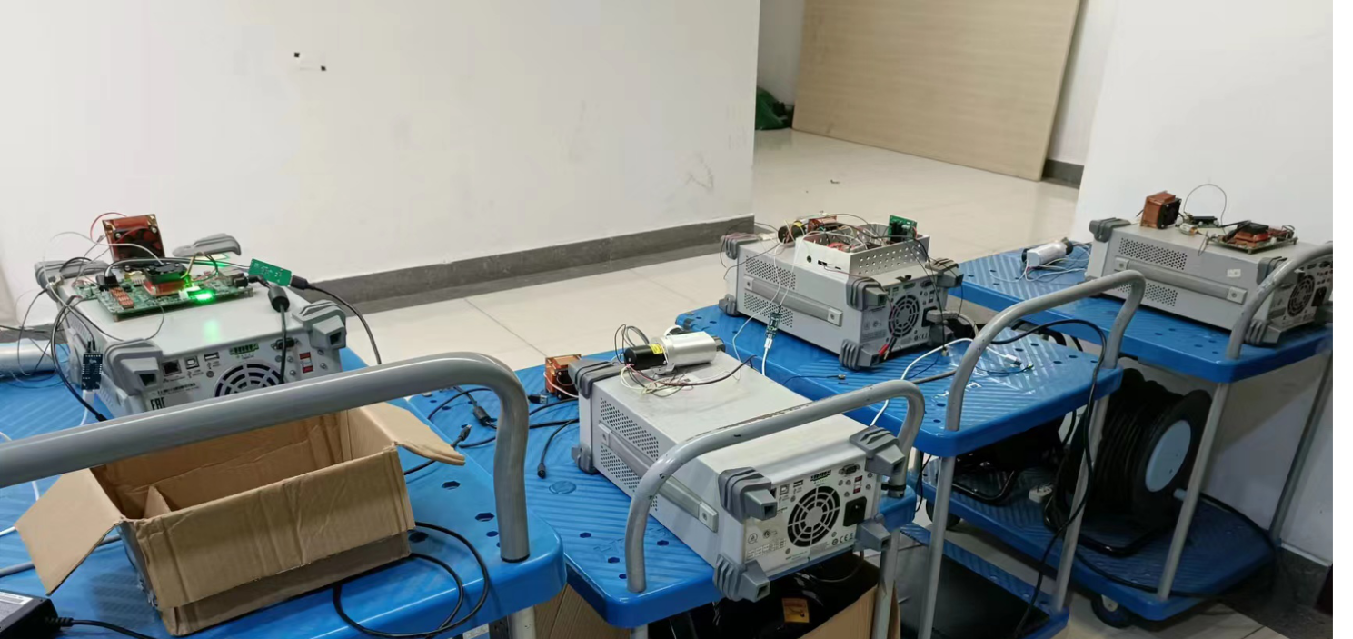}
	\caption{Lab test scenario.}
	\label{scenario_lab}
\end{figure}

\begin{figure}[htbp]
	\centering
	\includegraphics[width=0.4\textwidth]{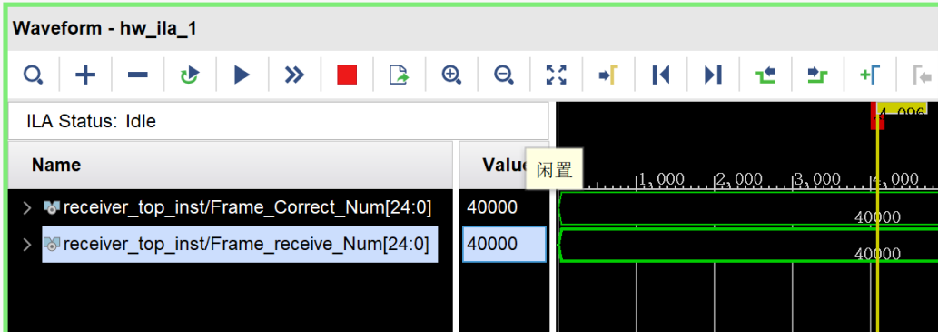}
	\caption{The experimental results of lab test.}
	\label{results_lab}
\end{figure}

We conduct real-time lab test on the system performance with the designed system parameters from the experimental results in Table \ref{slot_len} and the devices specified in Table \ref{system_parameter}. The experiment scenario is shown in Fig. \ref{scenario_lab}. We consider four communication nodes, consisting of one master node and three slave nodes. All nodes are put towards the white wall of lab. UV light from the LEDs is reflected off the white wall to the received area of the PMTs. Each node transmits $10000$ frames to the other nodes. The experimental results are shown in Fig. \ref{results_lab}. "Frame\_corret\_num" and "Frame\_receive\_num" denote the number of successfully decoded frames and the number of received frames, respectively. It can be observed that $40000$ frames are received and decoded successfully by one node, which demonstrates that the UV network system can work under lab scenario.

\begin{table}[htbp]
	\begin{center}
		\caption{Specification of Experimental Equipments and Devices}
		\label{system_parameter}
		\begin{tabular}{| c | c | c |}
			\hline
			\multirow{3}{4em}{UV LED}& Wavelength& $266$nm\\
			\cline{2-3}
			& Electric power& $58$W\\
			\cline{2-3}
			& Beam divergence& $120^\circ$\\
			\hline
			\multirow{4}{4em}{UV optical filter}& Peak wavelength& $264$nm\\
			\cline{2-3}
			& Peak transmission& $28.2\%$\\
			\cline{2-3}
			& Full width at half maximum& $20$nm\\
			\cline{2-3}
			& Aperture size& $\Phi 31.5$mm $\times$ $28.3$mm\\
			\hline
			\multirow{6}{4em}{PMT}& Spectral response& From $160$nm to $320$nm\\ 
			\cline{2-3}
			& Quantum efficiency& Around $30\%$ at $264$nm\\
			\cline{2-3}
			& Dark counts& $\leq 10$ per second\\
			\cline{2-3}
			& Anode pulse rise time& $2.2$ns\\
			\cline{2-3}
			& Detection bandwidth& $\geq 200$MHz\\
			\cline{2-3}
			& Aperture size& $\Phi 32.2$mm $\times$ $94$mm\\
			\hline
		\end{tabular}
	\end{center}
\end{table}

\subsection{Transmitter-receiver Field of View (FOV) Test with Experimental Results}\label{FOV}

\begin{figure}[htbp]
	\centering
	\includegraphics[width=0.4\textwidth]{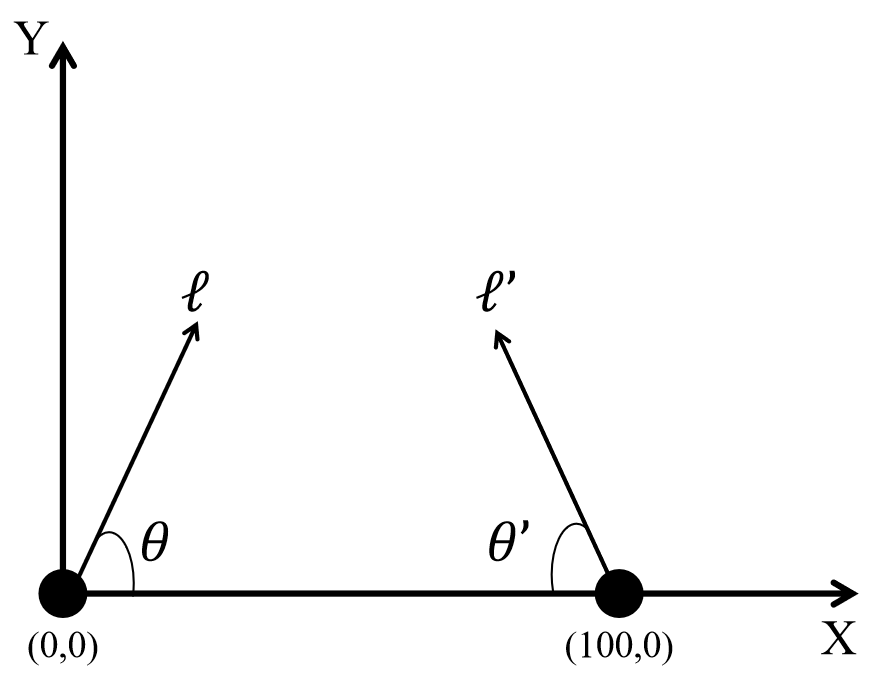}
	\caption{FOV test scenario.}
	\label{FOV_scenario}
\end{figure}

\begin{figure}[htbp]
	\centering
	\includegraphics[width=0.45\textwidth]{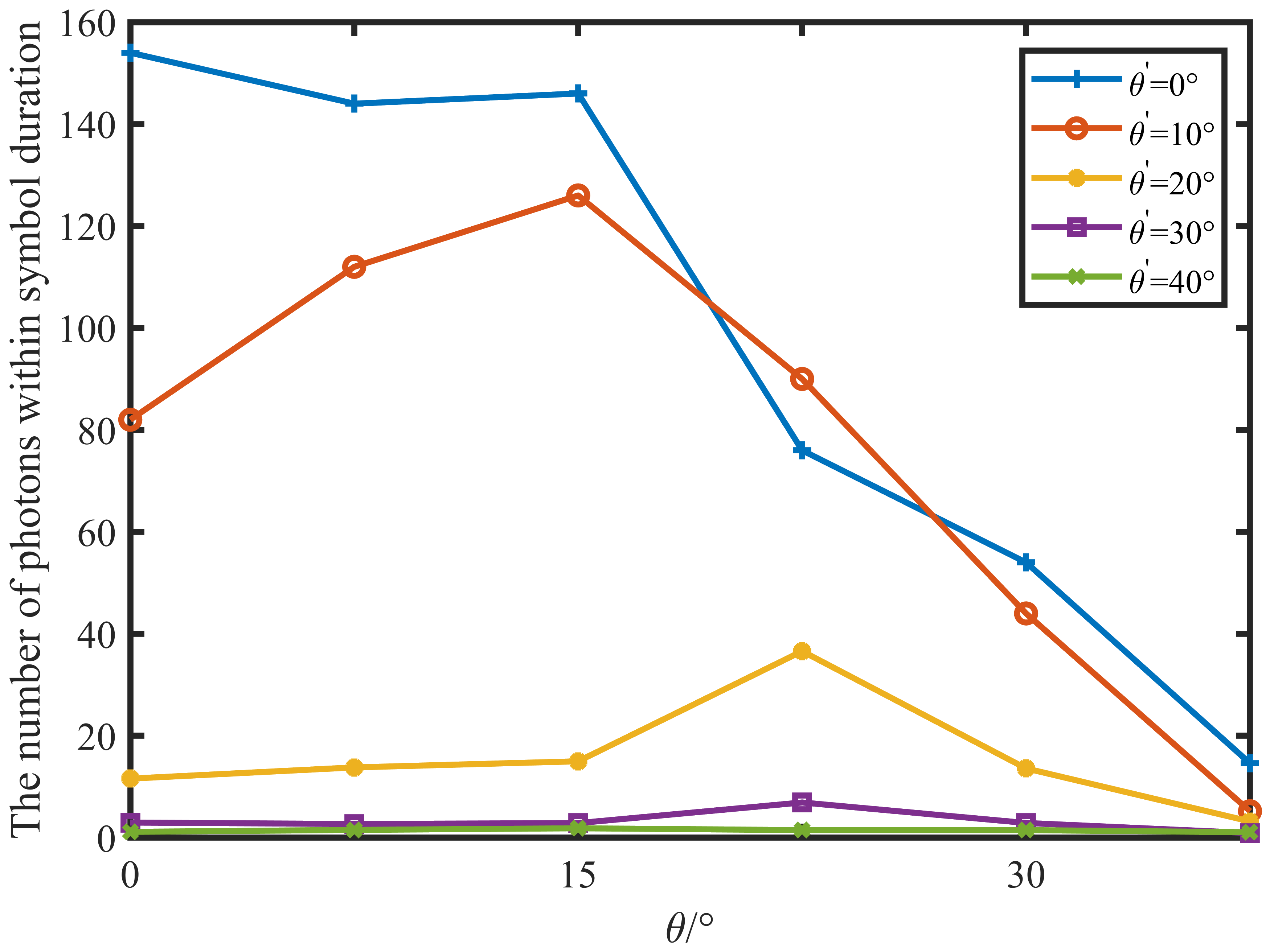}
	\caption{NLOS channel link gain with different rotations of LED and PMT.}
	\label{lg}
\end{figure}

We conduct transmitter-receiver field of view test on the NLOS channel link gain with the specified parameters in Table \ref{slot_len} and devices in Table \ref{system_parameter}  before outdoor real-time field test to determine the quantity of PMTs for one node $K$. The geometric configuration of the transmitter and receiver is shown in Fig. \ref{FOV_scenario}. The transmitter position is set to be ($0,0$) and the receiver position is set to be ($0,100$). $l$ and $l^{'}$ denote the orientation of the LED and PMT, respectively. Moreover, $\theta$ and ${\theta}^{'}$ denote the rotation angle of the LED and PMT. We achieve the NLOS channel link gain with $\theta$ from $0^{\circ}$ to $75^{\circ}$ and ${\theta}^{'}$ from $0^{\circ}$ to $40^{\circ}$. The results are shown in Fig. \ref{lg}. It is observed that Nlos channel link gain is significantly sensitive to the rotation of PMT due to the narrow FOV of PMT. The NLOS channel link gain dramatically drops below $4$ while $\theta^{'} = 30^\circ$, where the bit error rate (BER) of the communication link can not meet communication requirements. 

\subsection{Field Test over $110 \times 90 m^2$}

\begin{figure}[htp]
	\centering
	\includegraphics[width=0.45\textwidth]{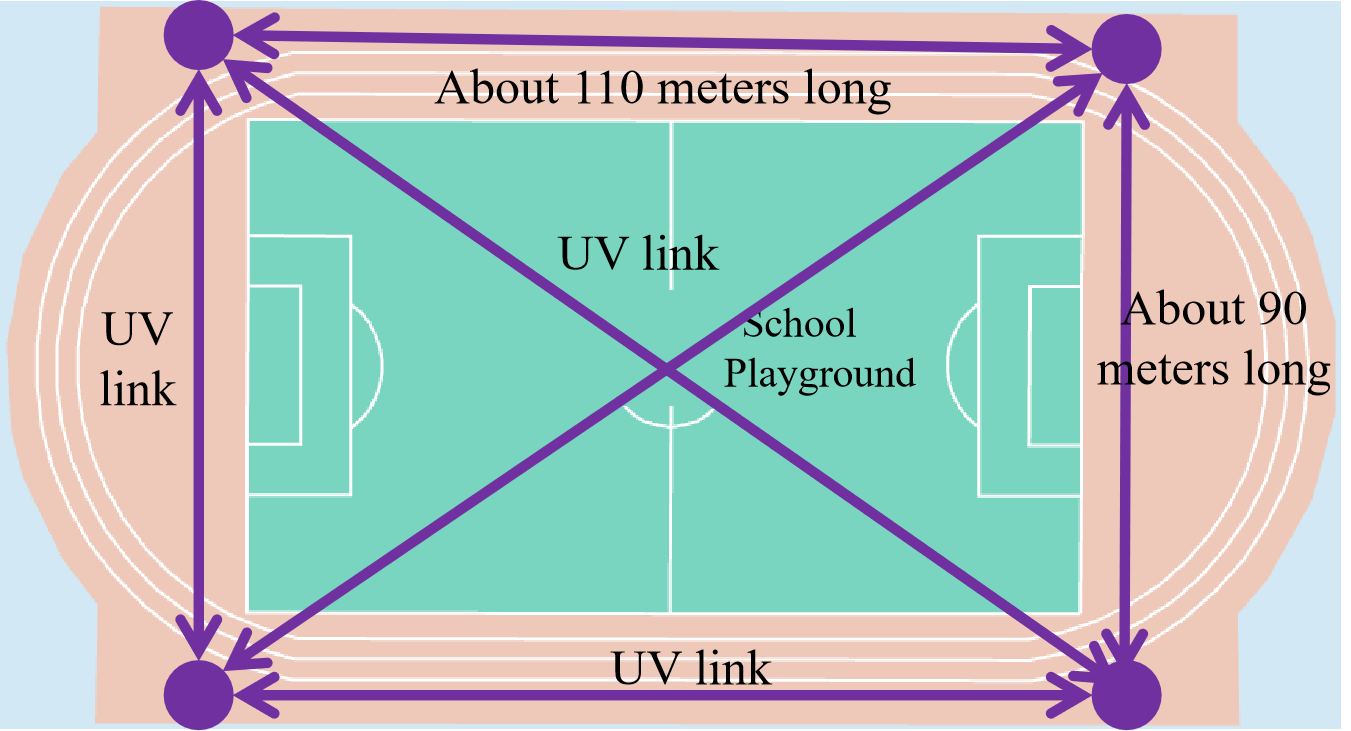}
	\caption{The network system layout.}
	\label{layout}
\end{figure}

\begin{figure}[htp]
	\centering
	\includegraphics[width=0.45\textwidth]{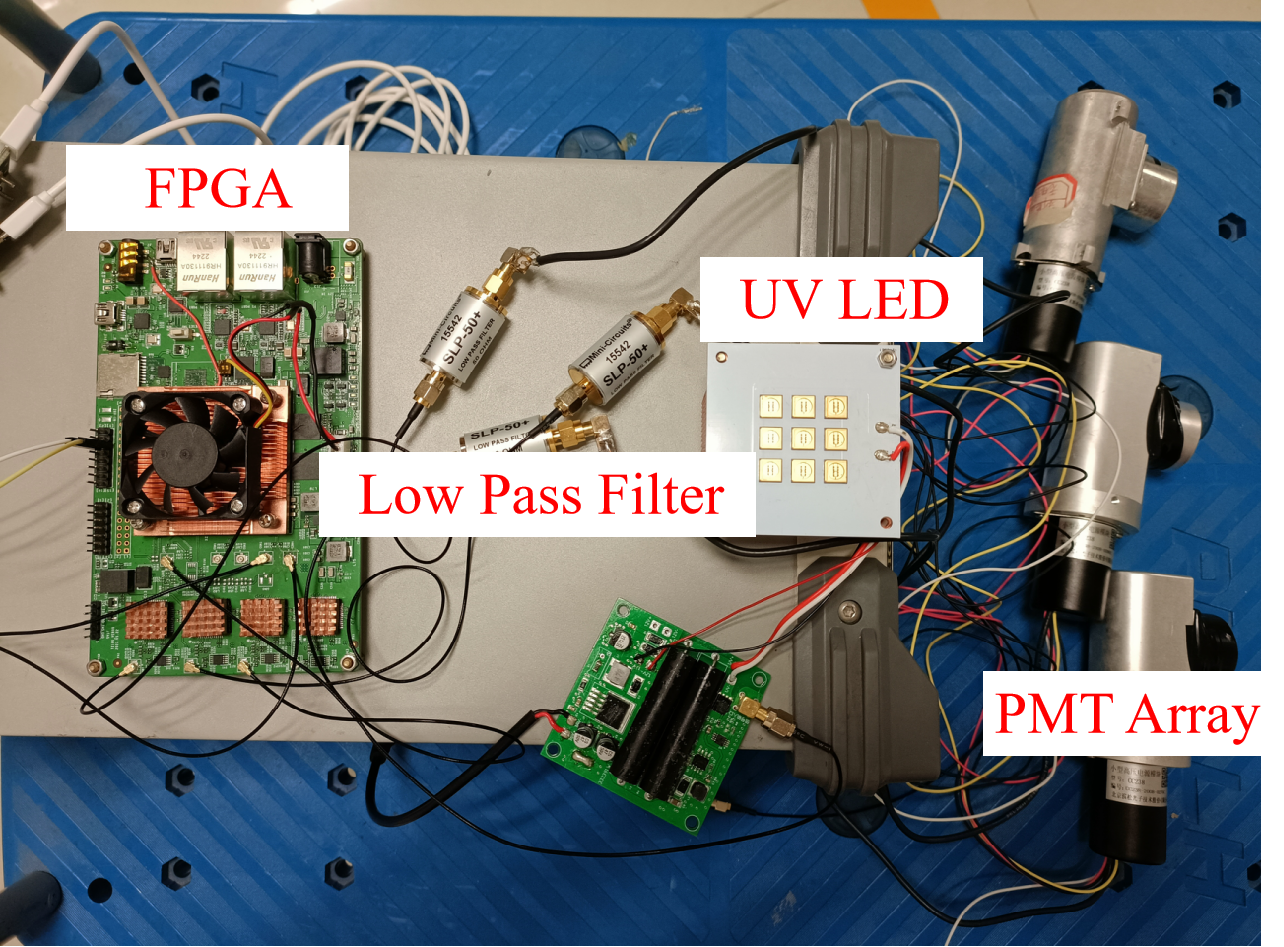}
	\caption{The test bed of each node.}
	\label{test_bed}
\end{figure}

\begin{figure}[htp]
	\centering
	\includegraphics[width=0.45\textwidth]{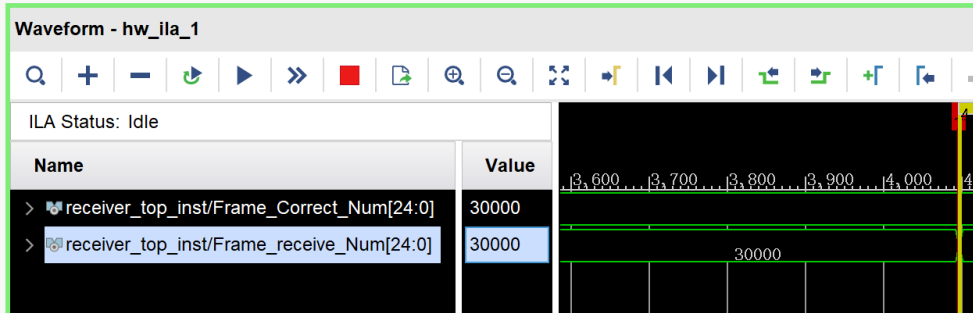}
	\caption{The experimental results of field test.}
	\label{results_field}
\end{figure}

Based on above experimental results in \ref{FOV}, we adopt the number of PMTs $K=3$ while the number of nodes in the network $N=4$. We carried out the outdoor field test on the real-time communication network consisting of one master node and three slave nodes. The network system layout is shown in Fig. \ref{layout}. The location of the system was placed in school playground. The date of outdoor field test was Oct.12th in 2023, where the weather conditions were 14$^\circ$C-23$^\circ$C, cloudy, and eastern wind of $2$ m/s. The four nodes are located at the four vertices of the rectangle. The width of school playground is about $90$ meters while the length of the rectangle is about $110$ meters. We show the hardware test bed of each node in Fig. \ref{test_bed}. The three PMTs of each node were placed towards two right-angle sides and diagonal line as in Fig. \ref{layout}, respectively. Each node transmits $10000$ frames. We adopt the ila core in the Vivado to obtain the results. The experimental results are shown in Fig. \ref{results_field}, where "Frame\_correct\_num" and "Frame\_receive\_num" denote the number of successfully decoded frames and the number of received frames, respectively. It can be observed that each node can successfully decode all the transmitted frames from the other nodes.

\section{Conclusion}\label{conclusion}

We have designed a NLOS UV communication network system, where a beacon-enabled TDMA scheme is adopted. We have also analyzed the existing time delays of the time sychronization and designed algorithms for beacon transmission, time compensation and time slot transition. We have conducted experiments to evaluate the time sychronizaiton error between nodes and the network system is specified based on the experimental results. We have also verified that the network system can work basd on the lab test. Furthermore, we have conducted outdoor FOV test to evaluate the FOV of the LEDs and PMTs. Based on the experimental results, we have finished the outdoor field test on the school playground with the transmission range over $110\times90$m$^2$, where the system throughput reaches $800$kbps.

\bibliographystyle{IEEEtran}
\bibliography{Beacon_TDMA_UV_Network}

\end{document}